\documentstyle[rotate,epsf,aps,twocolumn]{revtex}
\begin{document}
\newcommand{\be}{\begin{equation}}
\newcommand{\en}{\end{equation}}
\newcommand{\bea}{\begin{eqnarray}}
\newcommand{\ena}{\end{eqnarray}}

\draft

\title{Traveling time and traveling length in critical percolation
clusters}

\author{Youngki Lee,$^1$
Jos\'e S. Andrade Jr.,$^2$
Sergey V. Buldyrev,$^1$
Nikolay V. Dokholyan,$^1$ \\
Shlomo Havlin,$^3$ 
Peter R. King,$^4$
Gerald Paul,$^1$ and
H. Eugene Stanley$^1$
} 

\address{
$^1$Center for Polymer Studies and Department of Physics,
Boston University, Boston, Massachusetts 02215 \\
$^2$Departamento de F\'{\i}sica, Universidade Federal do Cear\'a,
    60451-970 Fortaleza, Cear\'a, Brazil \\
$^3$Minerva Center and Department of Physics, Bar-Ilan University, Ramat  
    Gan, Israel\\
$^4$BP Amoco Exploration, Sunbury-on-Thames, Middlesex, TW16 7LN, United
Kingdom and Department of Engineering, Cambridge University, Cambridge,
United Kingdom\\
}

\date{Last modified: Aug 18, 1999; Printed: \today}

\maketitle

\medskip

\begin{abstract}

We study traveling time and traveling length for tracer dispersion
in two-dimensional bond percolation, modeling flow by tracer particles
driven by a pressure difference between two points separated
by Euclidean distance $r$.
We find that the minimal traveling time $t_{min}$ scales as
$t_{min} \sim r^{1.33}$,
which is different from the scaling of the most probable
traveling time, ${\tilde t} \sim r^{1.64}$.
We also calculate the length of the path corresponding to the
minimal traveling time and find $\ell_{min} \sim r^{1.13}$
and that the most probable traveling length scales
as ${\tilde \ell} \sim r^{1.21}$.
We present the relevant distribution functions and scaling relations.

\end{abstract}



The study of flow in porous media has many applications, such as
hydrocarbon recovery and ground-water pollution
\cite{sahimi95,saffman,guyon,koplik1,bunde96}.  Here we study an
incompressible flow on two-dimensional bond percolation clusters
\cite{comment_a} at criticality where fluid is injected at point $A$ and
recovered at point $B$ separated from point $A$ by Euclidean distance
$r$.  At time $t=0$ we add a passive tracer \cite{comment_b} at the
injection point \cite{comment_bc}.  We investigate the scaling
properties of the distributions of {\it traveling time, traveling
length, minimal traveling time, and the length of the path corresponding
to the minimal traveling time} of the tracer particles.  We find new
dynamical scaling exponents associated with these distributions.

Our first step is to calculate the pressure difference across each bond
by solving Kirchhoff's law, which is equivalent to solving the Laplace
equation.  The velocity across a given bond is proportional to the
pressure difference across the bond; we normalize the velocities
assuming the total flow between $A$ and $B$ is fixed, independent of the
distance between $A$ and $B$ and the realization of the porous media
\cite{comment_c}.

We simulate the flow of tracers using a particle-launching algorithm
(PLA) \cite{comment_d}, where a tracer particle starting from the
injection point $A$ travels through the medium along a path connected to
recovery point $B$ \cite{comment_e}. The probability $p_{ij}$ that
a tracer particle at node $i$ selects an outgoing bond $(ij)$ is
proportional to the velocity of flow on that bond; $p_{ij}=v_{ij} / \sum_{k}
v_{ik}$, where the $k$ summation should be taken over all outgoing bonds,
i.e., for $v_{ik}>0$.  In this process, the time taken to pass through
the bond $(ij)$ is inversely proportional to the velocity of that bond,
i.e., $t_{ij}=1/v_{ij}$.

We measure the distributions, $P({\tilde t})$ and $P({\tilde \ell})$,
of the traveling time ${\tilde t}$ and
the traveling length ${\tilde \ell}$ between $A$ and $B$
for $10 000$ tracer particles for each realization.
We sample over $10 000$ different realizations with the two points $A$
and $B$ fixed. For each realization, we also find the minimal traveling
time and the path which corresponds to the minimal traveling time to
obtain $P(t_{min})$ and $P(\ell_{min})$.
We run the simulation for system size $L\times L$ where $L=1000 \gg r$,
and find a well-defined region where the distributions
follow the scaling form \cite{havlin87}
\be
P(x)= A_x\left({x \over x^*}\right)^{-g_x} f\left({x \over x^*}\right)
\label{eq:dist}
\en
where $x$ denotes ${\ell}_{min}$, $t_{min}$, $\tilde{\ell}$ or
$\tilde{t}$.
The normalization constant is given by $A_x \sim (x^*)^{-1}$
and we find the scaling functions to be of the form
$f(y)=\exp(-a_xy^{-\phi_x})$.
The maximum of the probability is at $x^*$.
Simulation shows that $x^*$ has a power-law dependence
on the distance $r$,
\be
x^* \sim r^{d_x}.
\en
The exponents $\phi_x$ and $d_x$ are related by $\phi_x=1/(d_x-1)$
\cite{degennes79}.
The scaling function $f$ decreases sharply when
$x$ is smaller than $x^*$. The lower cutoff
is due to the fact that the traveling distance cannot be
smaller than the distance $r$.

The path which takes minimal time is not always the shortest
path. However we find that the distribution of $\ell_{min}$ coincides
with the distribution of the chemical lengths between points separated
by distance $r$ studied in detail in Ref.~\cite{dokholyan98}.

In Figs.~$1(a)$, $2(a)$, and $3(a)$, we show the log-log plots of
distributions $P(t_{min})$, $P(\tilde{\ell})$, and $P(\tilde{t})$,
respectively.  For different distances $r=4,8,16,32,64$, and $128$, we
determine the characteristic size $x^*$ as the peak of the distribution.
In Figs. $1b$, $2b$, and $3b$, we plot $x^*$ versus distance $r$ in
double logarithmic scale and linear fitting yields the exponents $d_x$
for each distribution. In Figs.~$1(c)$, $2(c)$, and $3(c)$ we collapse
the data by rescaling $x$ by its characteristic size $x^*$. All
distributions are consistent with the scaling form of
Eq. (\ref{eq:dist}). The measured values of scaling exponents are
summarized in Table I.

\begin{figure}[htb]
\centerline{ \epsfxsize=5.0cm
\rotate[r]{ \epsfbox{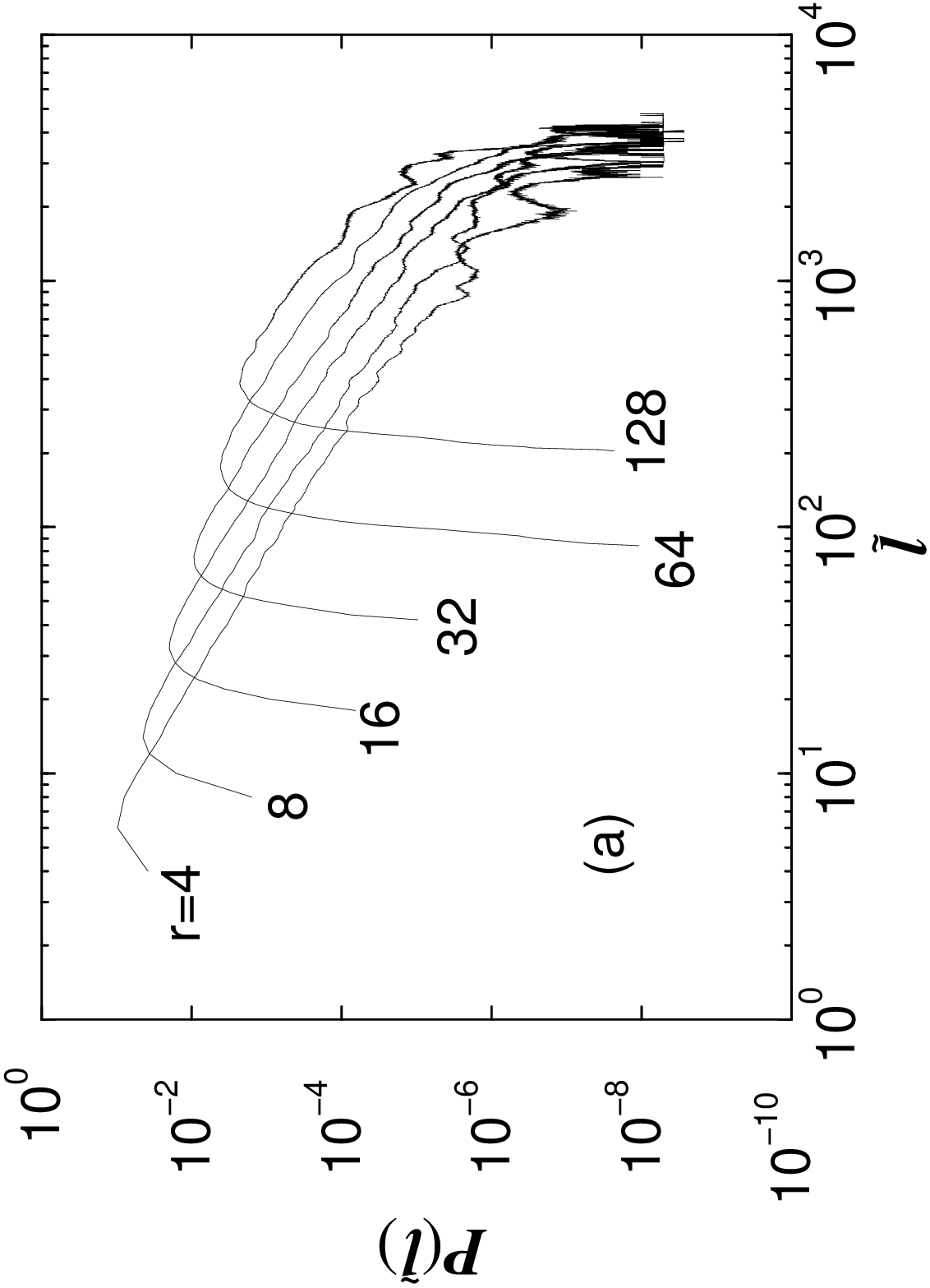} }
\vspace*{1.0cm}
}
\centerline{ \epsfxsize=5.0cm
\rotate[r]{ \epsfbox{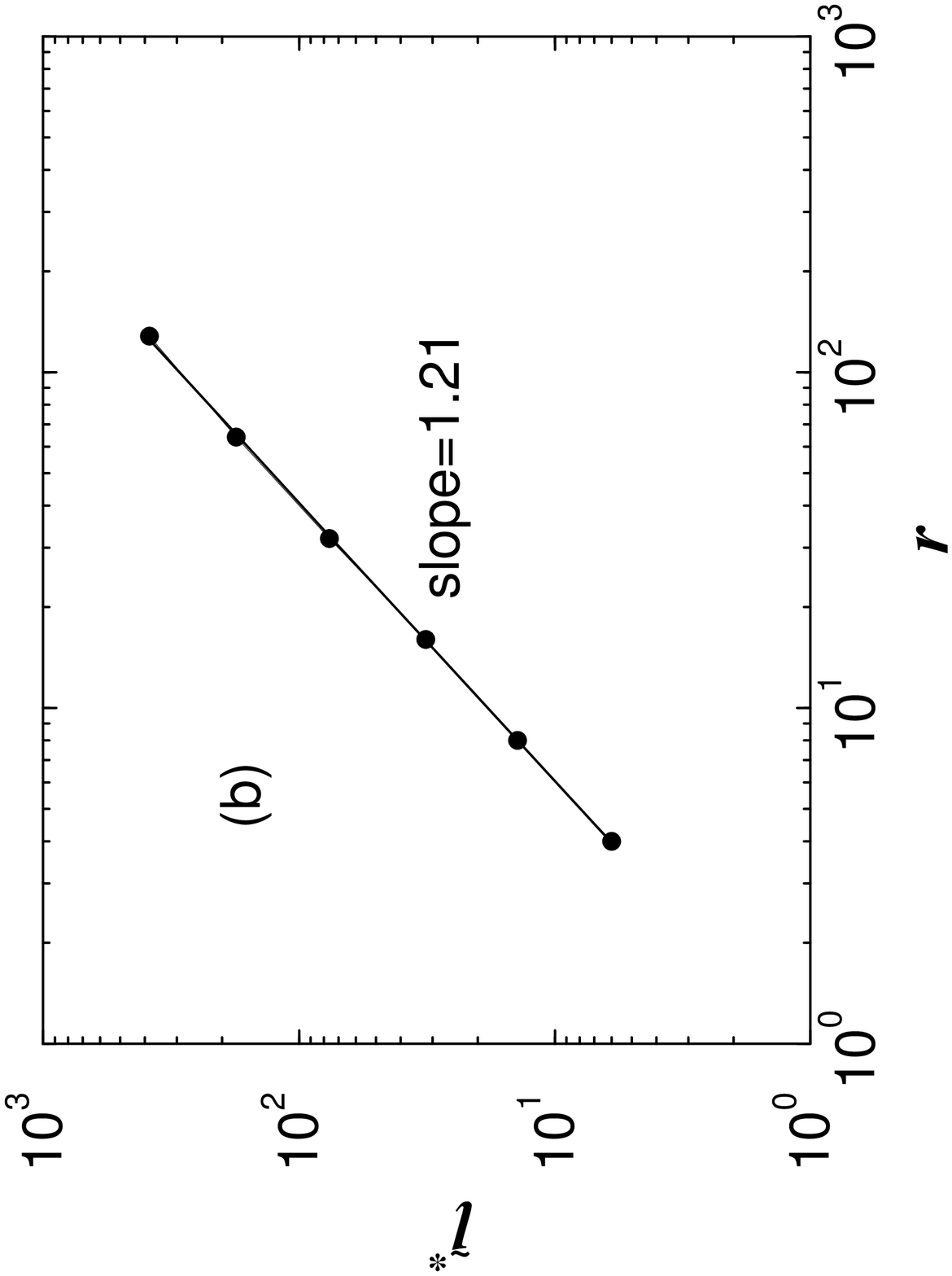} }
\vspace*{1.0cm}
}
\centerline{ \epsfxsize=5.0cm
\rotate[r]{ \epsfbox{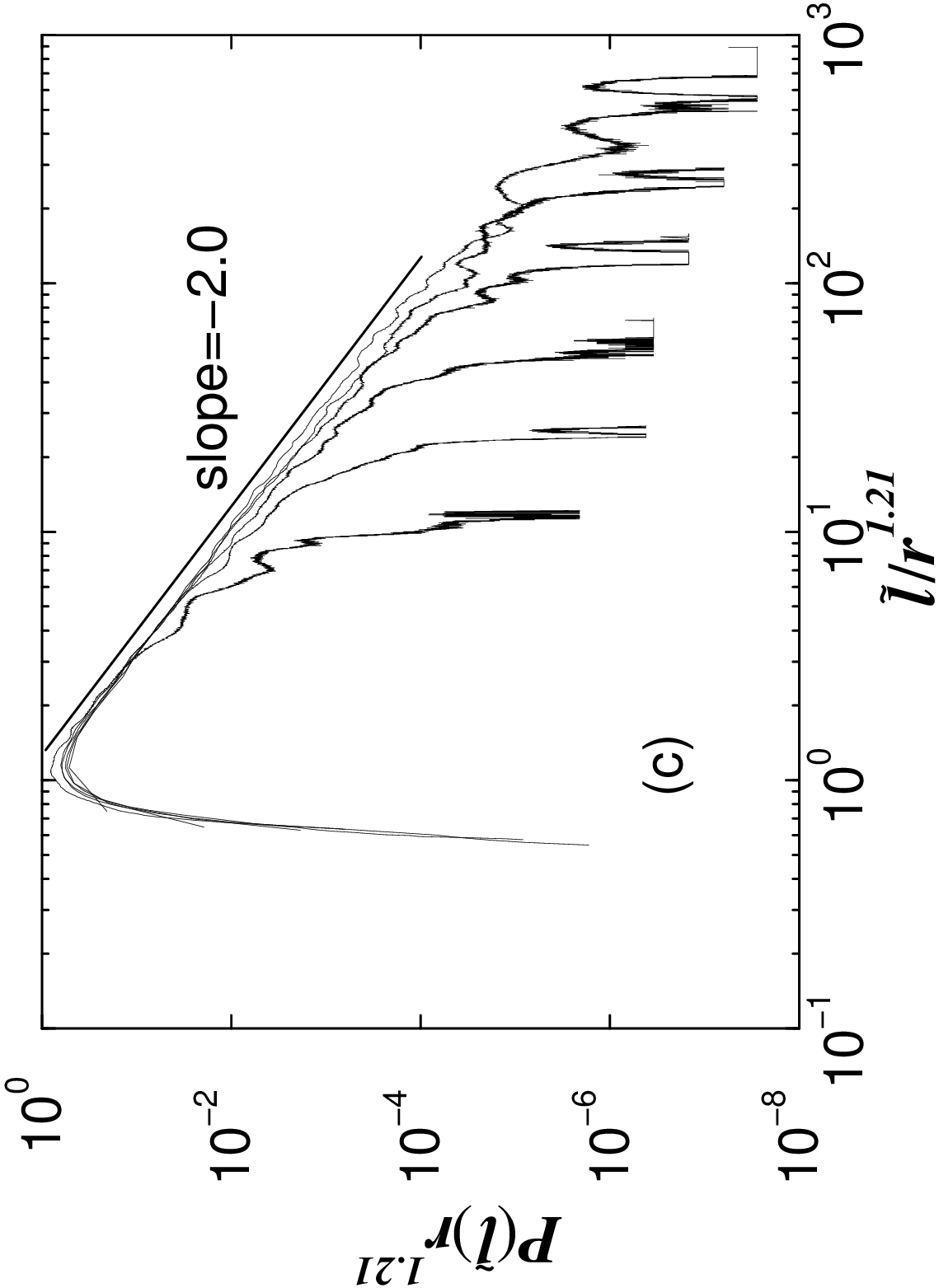} }
\vspace*{1.0cm}
}
\caption{
(a) Log-log plot of traveling distance distribution $P(\tilde{\ell})$
for $r=4,8,16,32,64$, and $128$.
(b) Log-log plot of the most probable traveling length
versus $r$. A linear fit yields $d_{\tilde{\ell}}=1.21 \pm 0.02$
(c) The data obtained by rescaling the traveling length with its
characteristic length $\tilde{\ell}^* \sim r^{1.21}$.
A fit of the power-law regime gives $g_{\tilde{\ell}}=2.0 \pm 0.05$.
}
\label{fig:trlength}
\end{figure}

As shown in Fig.~\ref{fig:trlength}(b), the most probable traveling
length $\tilde{\ell}^*$ scales as ${\tilde \ell}^* \sim
r^{d_{\tilde{\ell}}}$ where $d_{\tilde{\ell}}=1.21 \pm 0.02$. Note that
$d_{\tilde \ell}$ is significantly different from the minimal path
exponent $d_{min}=1.130 \pm 0.002$ \cite{herrmann88}, while it is within
the error bars of the exponent for the optimal path in random energy
landscapes, $d_{opt}=1.2 \pm 0.02$ \cite{cieplak94}, and the shortest
path in invasion percolation with trapping, $d_{opt}=1.22 \pm 0.01$
\cite{porto97}.

In many transport problems, the characteristic time scales with the
characteristic length with a power law, $t^* \sim (\ell^*)^z$. Since
$t^*$ scales as $r^{d_t}$ and $\ell^*$ scales as $r^{d_{\ell}}$, it is
reasonable to assume that $z=d_t/d_{\ell}$. Combining this relation,
the relation $t \sim \ell^z$, Eq.~(\ref{eq:dist}), and the identity
$P({\ell}_{min})d{\ell_{min}}=P(t_{min})dt_{min}$, we obtain scaling
relations between exponents,
\be
(g_{\ell_{min}}-1)d_{\ell_{min}}=(g_{t_{min}}-1)d_{t_{min}}
\label{eq:scaling}
\en
This scaling relation is well satisfied by the set of scaling exponents
given in Table I.

Because of flow conservation, the velocity at distance $r'$ from point
$A$ should scale inversely proportional to the number of bonds at this
distance, which scales as $(r')^{d_B -1}$ where $d_B$ is the fractal
dimension of the transport backbone.  Then the traveling time for a
particle to travel the distance $r$ is given by
\be
{\tilde t}^*(r) \sim \int_0^r {1 \over v(r')} dr' \sim r^{d_B}.
\en
Note that $\tilde{t}^*(r)$ is the most probable traveling time in our
system, so we obtain the scaling relation $d_{{\tilde t}}=d_B$. Thus,
the most probable traveling time is characterized by the transport
backbone dimension of the media. This result is consistent with the
homogeneous case, where $d_B=2$. The most recently reported value for
the fractal dimension of the backbone is $d_B=1.6432 \pm 0.0008$
\cite{grassberger99} for $d=2$, which is in agreement with our results
(Table I).

The minimal traveling time is the sum of inverse velocities over the
fastest path where as noted above the fastest path is statistically
identical to the shortest path.  While the velocity distribution has
been studied extensively \cite{arcangelis85} (e.g. it is known to be
multifractal), because the velocities along the path are correlated,
how the minimum traveling time distribution is related to the local
velocity distribution is an open challenge for further research.

\smallskip

We thank A. Coniglio, D. Stauffer, and especially M. Barth\'{e}l\'{e}my
for fruitful discussions, and BP Amoco for financial support. We also
thank J. Koplik and S. Redner for discussions concerning the limitation
of a PLA.


\newpage

\begin{center}
\begin{tabular}{ccc}\hline \hline
$x          $&~~~~$d_x          $&~~~~$g_x          $     \\\hline
$\ell_{min}  $&~~~~$1.13 \pm 0.01$&~~~~$2.14 \pm 0.05$     \\
$t_{min}     $&~~~~$1.33 \pm 0.05$&~~~~$1.90 \pm 0.05$     \\
$\tilde{\ell}$&~~~~$1.21 \pm 0.02$&~~~~$2.00 \pm 0.05$     \\
$\tilde{t}   $&~~~~$1.64 \pm 0.02$&~~~~$1.62 \pm 0.05$     \\\hline \hline
\end{tabular}
\label{tab:exponents}
\\
\vspace{0.5cm}
\end{center}

Table I. Results for the exponents. Our $d=2$ results for
$d_{\ell_{min}}$ and $g_{\ell_{min}}$ are within error bars of $d_{min}$
and $g'_\ell$ in Ref.~\cite{dokholyan98}. For comparison, the
theoretical values of $d_x$ and $g_x$ for $d=6$ are all $2$.

\begin{figure}[htb]
\centerline{ \epsfxsize=5.0cm
\rotate[r]{ \epsfbox{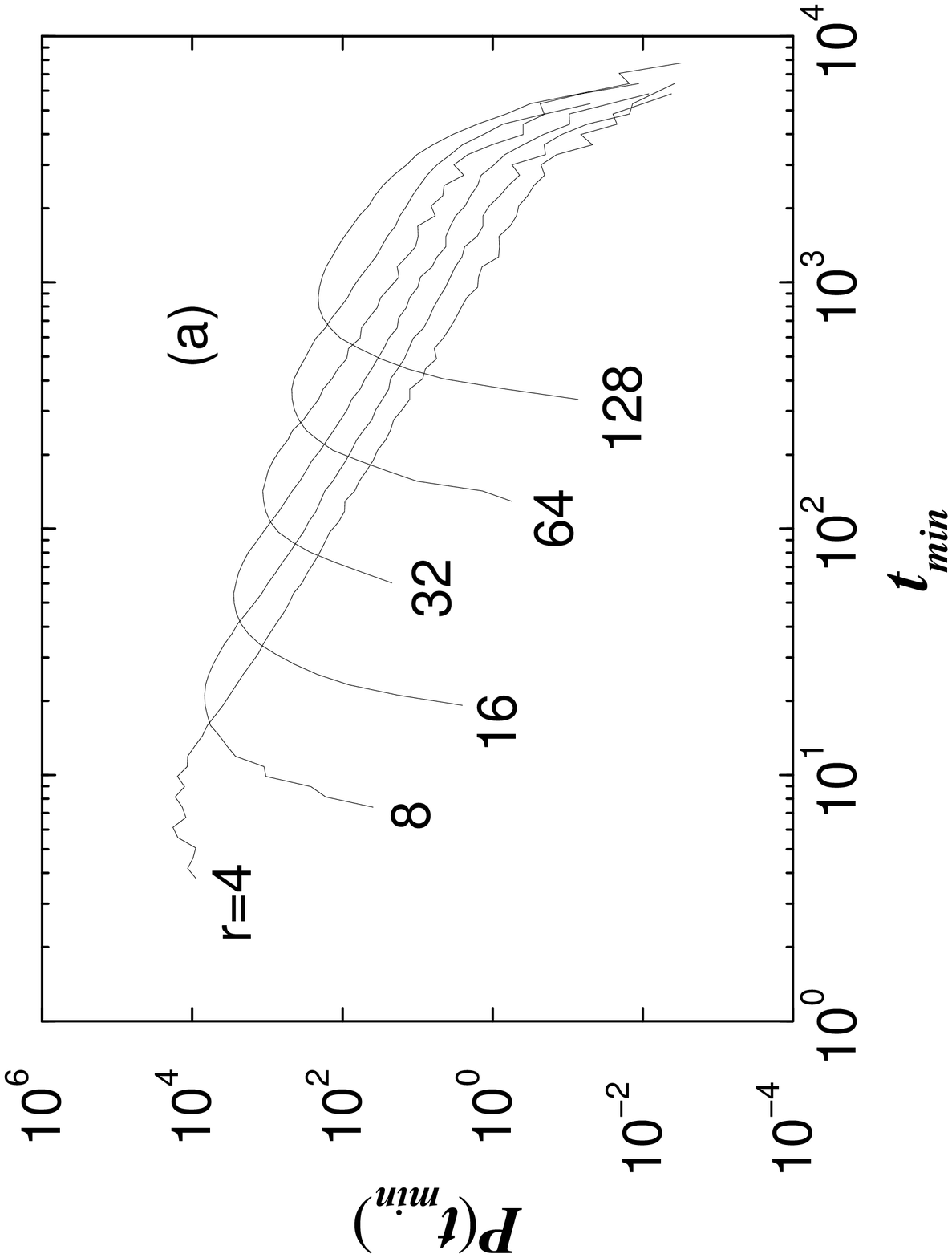} }
\vspace*{1.0cm}
}
\centerline{ \epsfxsize=5.0cm
\rotate[r]{ \epsfbox{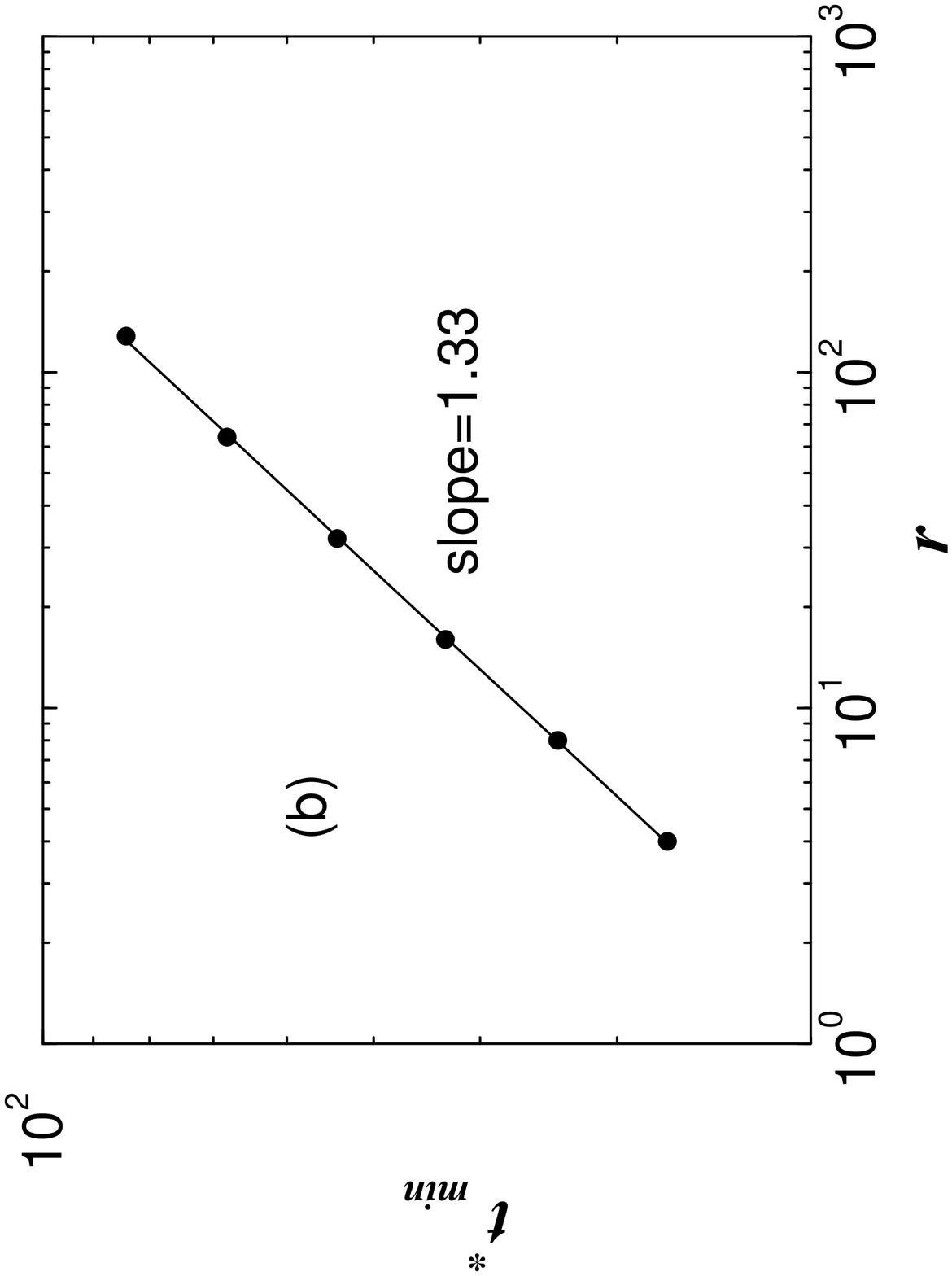} }
\vspace*{1.0cm}
}
\centerline{ \epsfxsize=5.0cm
\rotate[r]{ \epsfbox{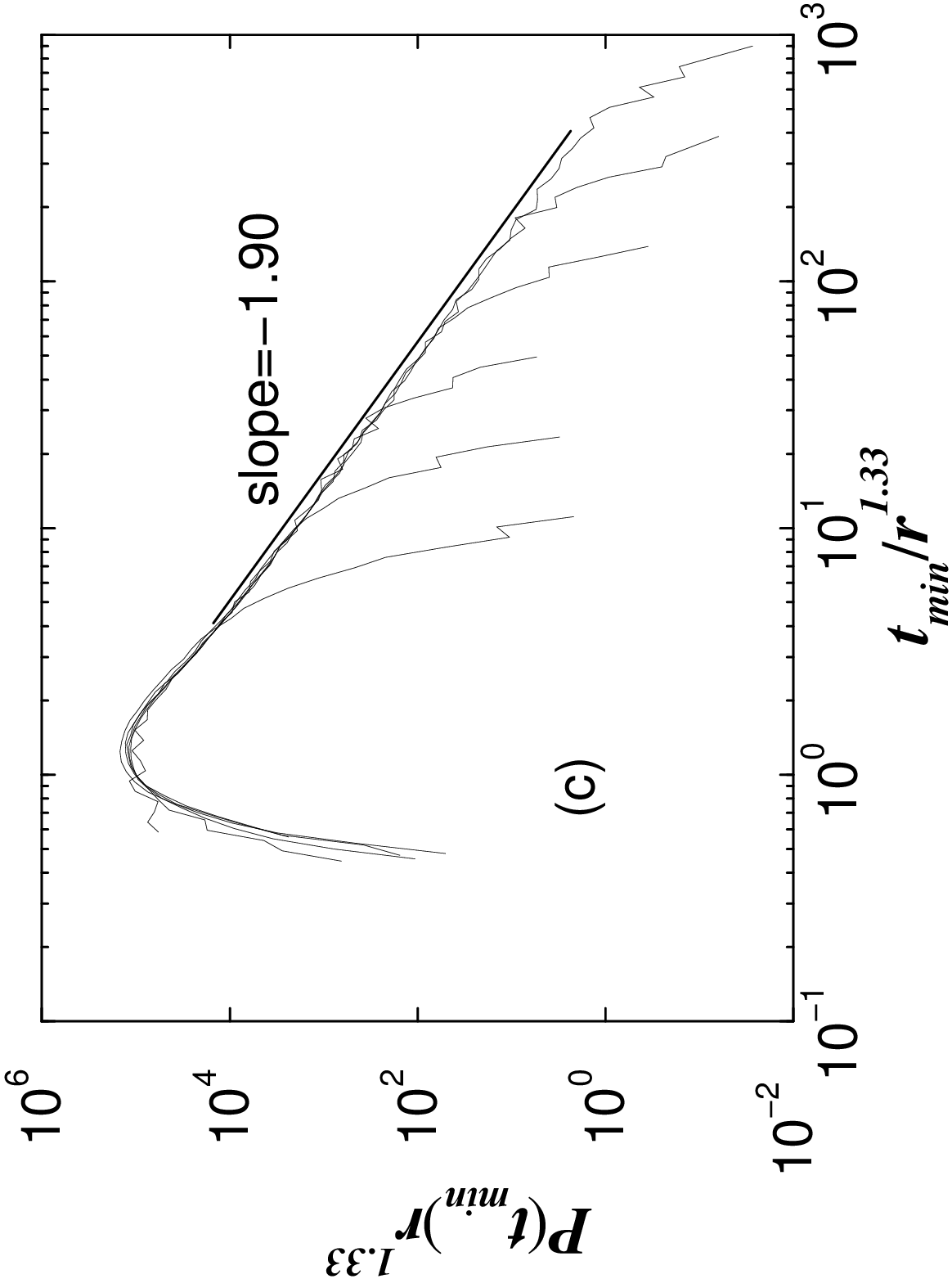} }
\vspace*{1.0cm}
}
\caption{
(a) Log-log plot of the minimal traveling time distribution $P(t_{min})$
for separations $r=4,8,16,32,64$, and $128$ between injection
and recovery points.
(b) Log-log plot of the most probable minimal traveling time
versus $r$. A linear fit yields $d_{t_{min}}=1.33 \pm 0.05$.
(c) The data obtained by rescaling the minimal time
with its characteristic time $t_{min}^* \sim r^{1.33}$.
A fit of the power-law regime gives $g_{t_{min}}=1.90 \pm 0.05$.
}
\label{fig:brtime}
\end{figure}

\begin{figure}[htb]
\centerline{ \epsfxsize=5.0cm
\rotate[r]{ \epsfbox{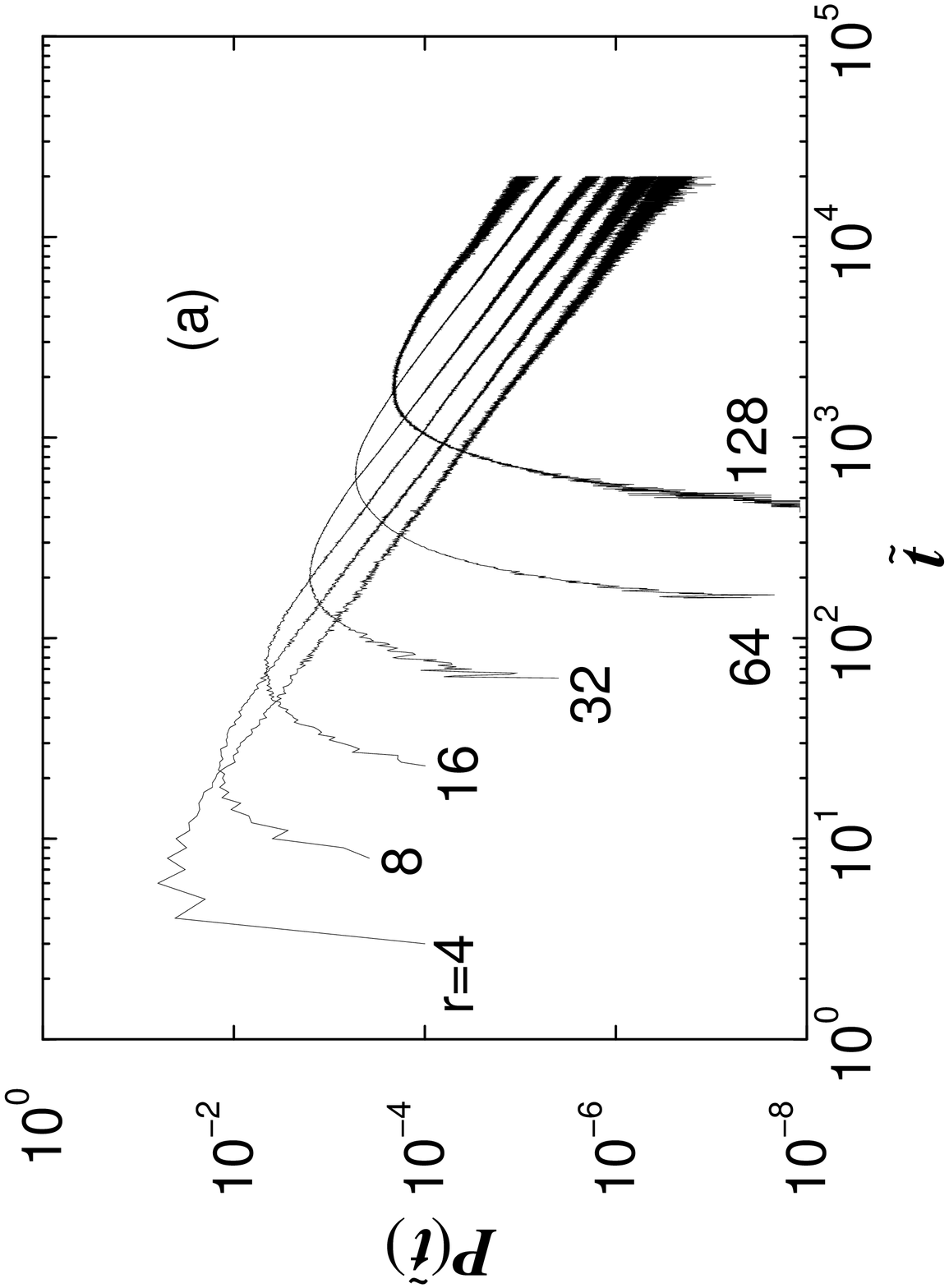} }
\vspace*{1.0cm}
}
\centerline{ \epsfxsize=5.0cm
\rotate[r]{ \epsfbox{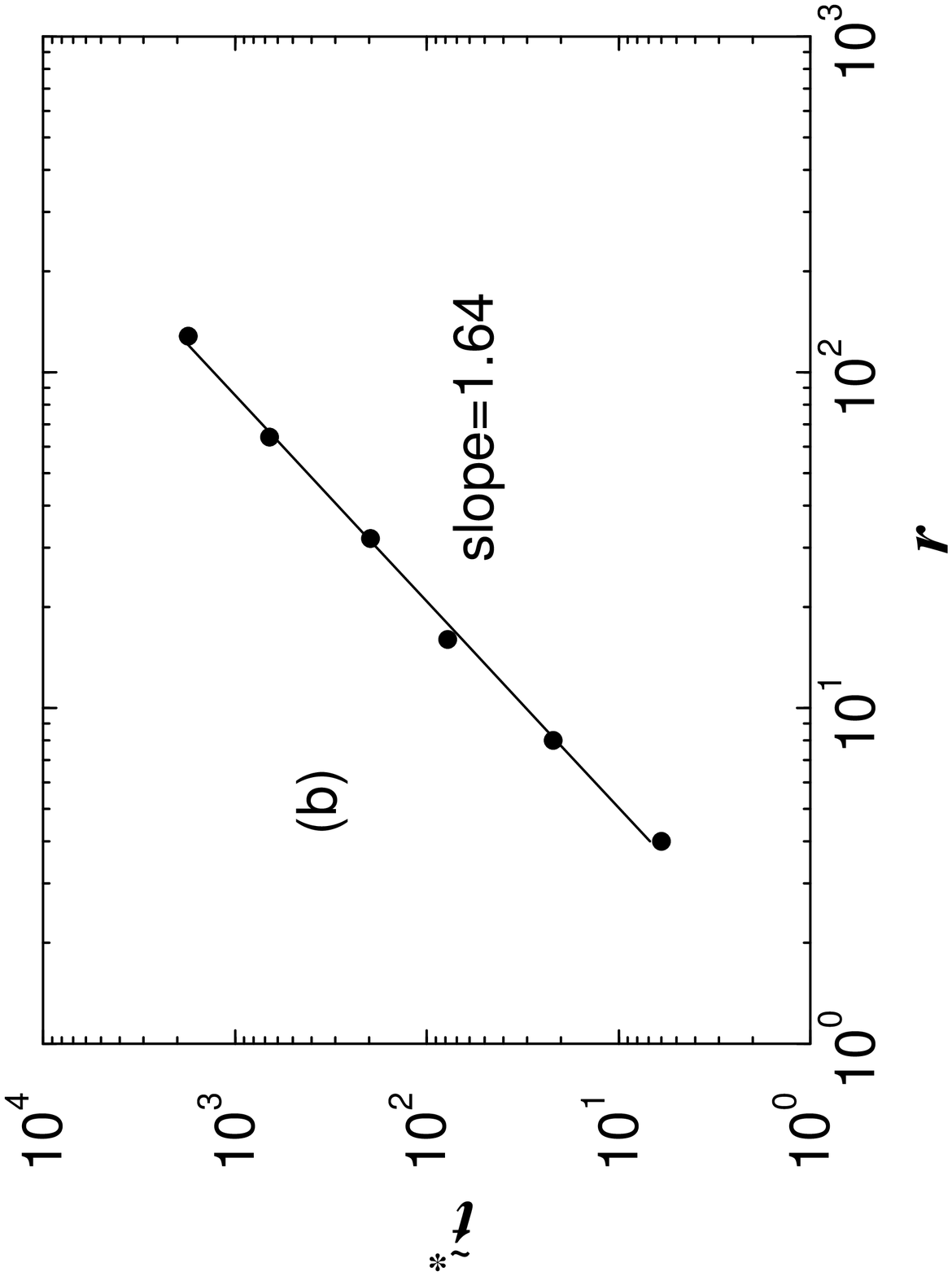} }
\vspace*{1.0cm}
}
\centerline{ \epsfxsize=5.0cm
\rotate[r]{ \epsfbox{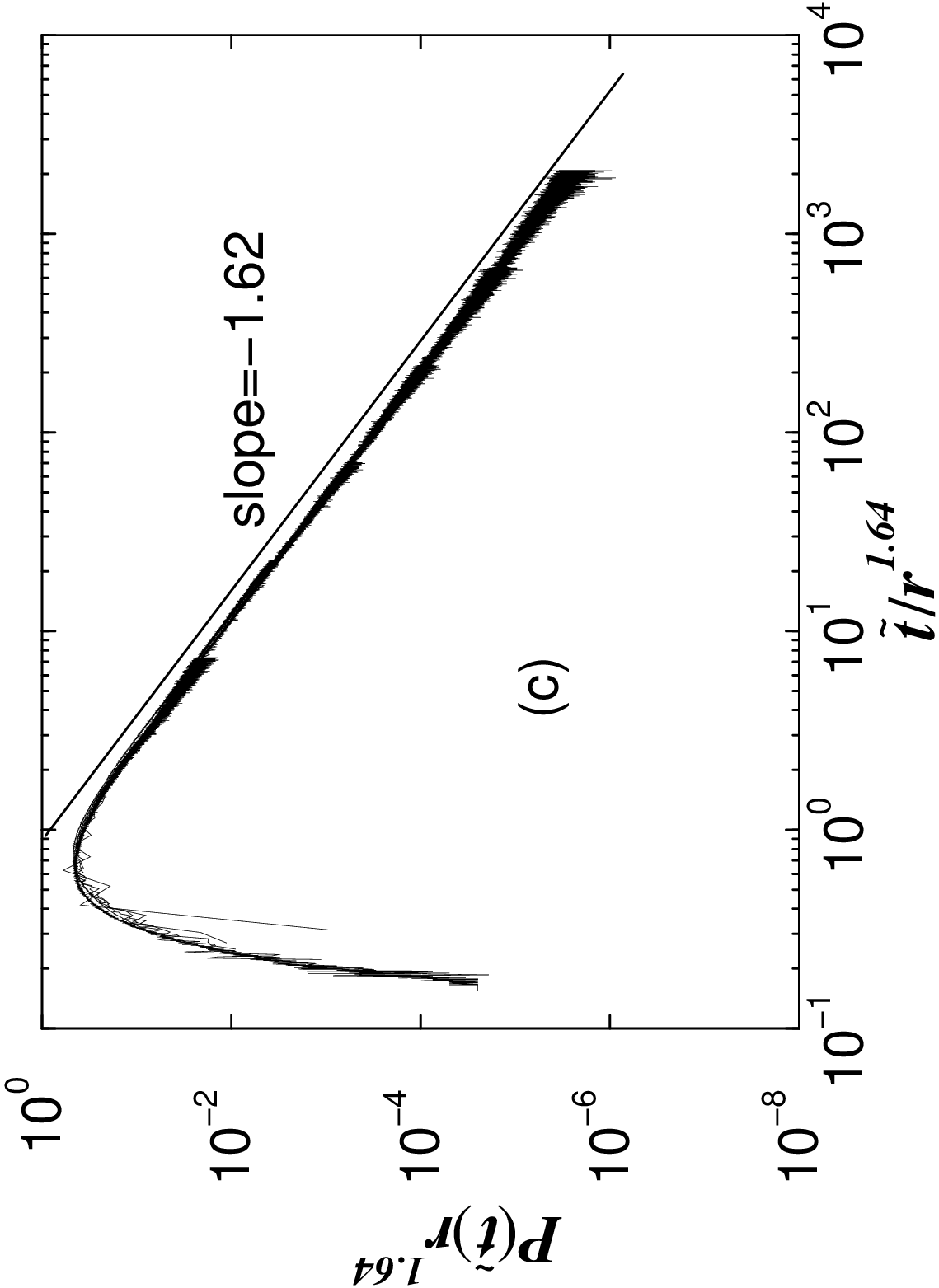} }
\vspace*{1.0cm}
}
\caption{
(a) Log-log plot of traveling time distribution $P(\tilde{t})$
for $r=4,8,16,32,64$, $128$. 
(b) Log-log plot of the most probable time
versus $r$. A linear fit yields $d_{\tilde t}=1.64 \pm 0.02$.
(c) The data obtained by rescaling the time with its characteristic time
${\tilde t}^* \sim r^{1.64}$. A linear fit of the power-law
regime gives $g_{\tilde t}=1.62 \pm 0.05$.
}
\label{fig:trtime}
\end{figure}


\begin{thebibliography}{99}
\bibitem{comment_a}
It is customary to model oil recovery for 
configurations in which the lateral extents are much greater
than the vertical extent using two-dimensional bond percolation.
\bibitem{comment_b}
By passive tracer we mean that the tracer moves only by convection,
ignoring molecular diffusion that is slow on the time scales
of interest, and that it is not absorbed by its surroundings.
\bibitem{comment_c}
This normalization realistically models oil recovery where constant
flow, as opposed to constant pressure, is maintained.
\bibitem{comment_d}
A PLA is a useful approximation for any flow whose velocity
field is known to be constant in time, but for which transport
properties cannot be solved either analytically or numerically.
A PLA is inaccurate for very large traveling times due to the low
probability of entering bonds with low potential differences
(and therefore long transit times; see also Ref. \cite{koplik1}).
The ``very long time regime'' where diffusion becomes
a significant factor is not considered in this paper.
\bibitem{comment_e}
The traveling path is randomly selected with a probability
out of all possible paths connecting two points.
If the path contains $n$ nodes, then the probability of the
path is determined as a product of the independent local
probabilities $p_{ij}$ over the number of nodes.
\end{thebibliography}
\end{document}